\documentclass[a4paper,11pt,twocolumn]{article}
\usepackage{cite}
\usepackage[T1]{fontenc}
\usepackage{graphicx}
\usepackage{authblk}
\usepackage{amsmath}
\usepackage[margin=0.8in]{geometry}

\usepackage{array}
\usepackage{caption}

\newcolumntype{M}[1]{>{\centering\arraybackslash}m{#1}}

\title{Supercontinuum generation in a graded-index multimode tellurite fiber}

\author[1,*]{Ekaterina Krutova}
\author[1]{Lauri Salmela}
\author[1]{Zahra Eslami}
\author[2,3]{Tanvi Karpate}
\author[3]{Mariusz Klimczak}
\author[2,3]{Ryszard Buczynski}
\author[1]{Go{\"e}ry Genty}

\affil[1]{Photonics Laboratory, Physics Unit, Tampere University, 33014 Tampere, Finland}
\affil[2]{Łukasiewicz Research Network – Institute of Microelectronics and Photonics, Al. Lotników 32/46, 02-668 Warsaw, Poland}
\affil[3]{Faculty of Physics, University of Warsaw, Pasteura 5, 02-093 Warsaw, Poland}

\affil[*]{Corresponding author: ekaterina.krutova@tuni.fi}

\begin{document}
\twocolumn[
\begin{@twocolumnfalse}
\maketitle
\begin{abstract}
We report the generation of a broadband supercontinuum (SC) from 790~nm to 2900~nm in a tellurite graded-index multimode fiber with a nanostructured core. We study the SC dynamics in different dispersion regimes and observe near-single mode spatial intensity distribution at high input energy values. Numerical simulations of the (3+1)D generalized nonlinear Schrödinger equation are in good agreement with our experiments. Our results open a new avenue for the generation of high-power mid-infrared SC sources in soft-glass fibers.
\end{abstract}
\end{@twocolumnfalse}
]



Complex nonlinear propagation dynamics in graded-index (GRIN) multimode fibers have attracted significant attention in the past few years \cite{article1, Saini:15, article2} in which a parabolic refractive index profile yields spatial modulation of the beam intensity along propagation. The resulting spatio-temporal coupling leads to specific nonlinear phenomena, including geometric parametric instabilities (GPI) \cite{gpi} and multimode soliton formation \cite{multimodesoliton}. Although GRIN multimode fibers can support a large number of modes, it has also been shown that the spatial intensity profile of the beam at the fiber output can exhibit a near-Gaussian distribution in the nonlinear regime under particular injection conditions due to beam self-cleaning dynamics \cite{krupanature}. These observations have opened up novel perspectives for generating high-power broadband supercontinuum (SC) with a near-Gaussian spatial intensity profile. 

Studies of nonlinear dynamics in GRIN fibers have been generally limited to commercial fibers made of silica with a transmission window of up to 2400~nm. Optical fibers made of soft glasses can however offer higher nonlinearity and a broader transmission range towards the infrared region \cite{scsoftglassreview}. While developing GRIN fibers operating in the mid-infrared region can be challenging, recently, a new manufacturing approach based on nanostructuring using multi-component oxide-based glasses has been introduced \cite{stackanddraw, article8}, allowing for the design and fabrication of GRIN multimode fibers made of lead-bismuth oxide glasses with extended transmission in the infrared \cite{Eslami}. 

Among soft glass materials, tellurite glasses possess a five times higher nonlinearity than lead-bismuth-gallate (PBG) oxide glasses and exhibit wider transmission in the mid-infrared \cite{telluritechar}. Compared to PBG, tellurite glasses can therefore allow for the generation of a SC with a larger bandwidth or require lower input power for a SC with comparable bandwidth. Furthermore, tellurite glasses possess superior chemical properties such as high thermomechanical and chemical stability, resistance to atmospheric humidity, and crystallization, making them attractive candidates for mid-infrared SC generation \cite{liao2009tellurite, shi2016multi, Saini:19}.

In this Letter, we report mid-infrared SC generation in a GRIN multimode tellurite fiber, showing that the concept of nanostructuring using multi-component glasses to design a GRIN multimode fiber can be applied to tellurite with an increased intrinsic nonlinearity. We demonstrate SC generation in both normal and anomalous dispersion regimes, with a SC spanning nearly two octaves in the anomalous dispersion regime. Additionally, characterization of the spatial intensity profile of the SC beam at the fiber output as a function of injected pulse energy shows signatures of beam self-cleaning dynamics. Our experimental results are confirmed by numerical simulations of the (3+1)D generalized nonlinear Schrödinger equation (GNLSE). 

\begin{figure}[t]
\centering
\includegraphics[width=\linewidth]{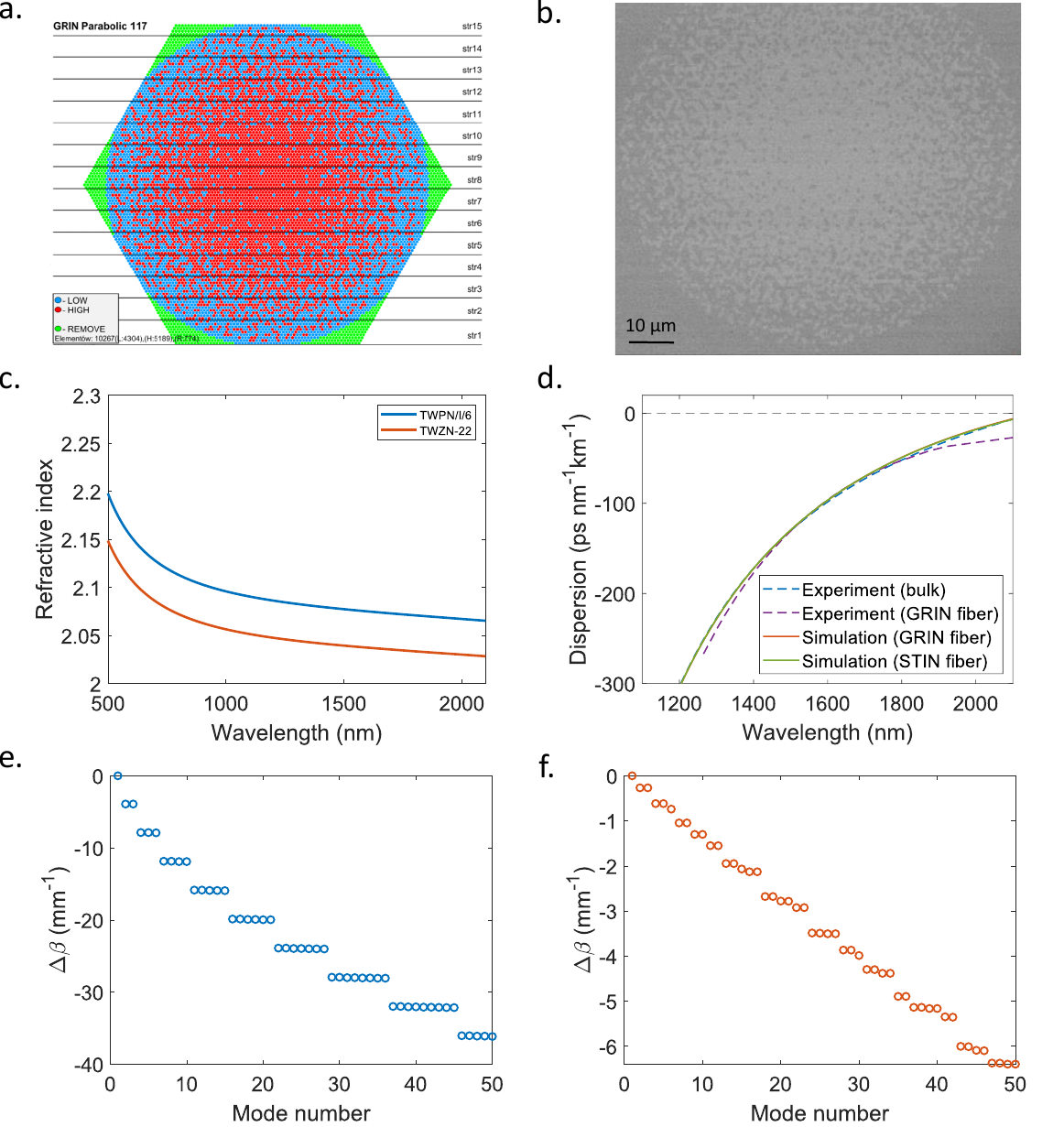}
\caption{a) Structural design of the GRIN multimode fiber; b) Scanning electron microscope image of the fabricated fiber cross-section; c) Refractive indices of two tellurite glasses TWPN/I/6 (red) and TWZN-22 (blue) as functions of wavelength; d) Experimentally measured dispersion of the tellurite TWPN/I/6 bulk glass (dashed blue) and GRIN tellurite MMF (dashed purple), numerically simulated for the GRIN fiber (red) and step-index fiber (green); e) Propagation constant relative to the fundamental mode for the first 50 modes of the GRIN tellurite fiber and f) propagation constant relative to the fundamental mode for the first 50 modes of a step-index tellurite fiber made of the same glasses.}
\label{fig:fiber_prop}
\end{figure}

\begin{figure}[b]
\centering
\includegraphics[scale=1]{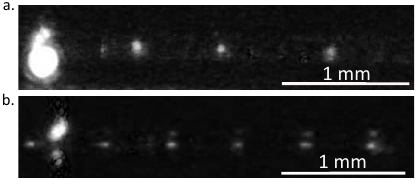}
\caption{ Experimentally observed spatial self-imaging in a) the tellurite GRIN MMF and b) a silica GRIN MMF (Thorlabs GIF50C) both excited by the residue of the OPA pump at $\lambda$ = 532~nm.}
\label{fig:ssi}
\end{figure}

In order to fabricate a tellurite fiber with a parabolic refractive index profile, a stack and draw approach similar to that in Ref.~\cite{Eslami} was used. In this approach, rods of glasses with close (but different) refractive indices are arranged according to a particular configuration yielding a quasi-continuous parabolic refractive index (see Supplementary Document for full fabrication details). Here, 10267 rods with 0.5 mm diameter made from two in-house developed tellurite glasses (denoted as TWPN/I/6 and TWZN-22) were used and their optimized arrangement is shown in Fig.~\ref{fig:fiber_prop}a. Before assembly, the glass rods are subjected to a sequence of cleaning, ionization, and evacuation procedures. Subsequently, the stacked rods are enclosed within a low-index tellurite glass tube (TWZN-22) and the preform is drawn into a thin fiber with a core size of 93 µm. A scanning electron microscope image of the resulting fiber structure is shown in Fig.~\ref{fig:fiber_prop}b. The refractive index of the two tellurite glasses modeled using a Sellmeier's equation is plotted in Fig.~\ref{fig:fiber_prop}c and the dispersion of the TWPN/I/6 bulk material and that of the fiber are shown in Fig.~\ref{fig:fiber_prop}d. The dispersion was measured by white-light interferometry based on a balanced Mach-Zehnder interferometer and fitting the data with a Laurent expansion. The numerically simulated dispersion profile of the fundamental mode is also plotted in the Figure for comparison. One can see that the dispersion is close to that of the bulk tellurite glass with a zero-dispersion wavelength (ZDW) at around 2200~nm, and also that the simulated dispersion of the fundamental mode of GRIN and step-index fibers are nearly identical. Additionally, we numerically simulated various higher-order modes supported by the fiber, and the relative difference between their propagation constant and that of the fundamental mode is illustrated in Fig.~\ref{fig:fiber_prop}e. For comparison, we also show in Fig.~\ref{fig:fiber_prop}f the relative propagation constant of the same set of modes assuming a step-index fiber made from the same tellurite glasses. Unlike the step-index fiber, the parabolic profile of the GRIN fiber yields an equidistant difference between the propagation constants of different mode groups. This feature leads to spatial self-imaging as observed in our experiments. A grayscale photograph of the self-imaging pattern is shown in Fig.~\ref{fig:ssi}a., confirming the parabolic index profile of the fabricated fiber. The self-imaging period is close to the theoretically calculated value $z_p = 0.69$~mm, however, one can also see that the period is not constant, indicative of geometrical inhomogeneity along the fiber. This is even more apparent when compared to the self-imaging observed in a commercial silica GRIN MMF (see Fig.~\ref{fig:ssi}b). A more detailed comparison can be found in the Supplementary document.

\begin{figure}[t]
\centering
\includegraphics[width=\linewidth]{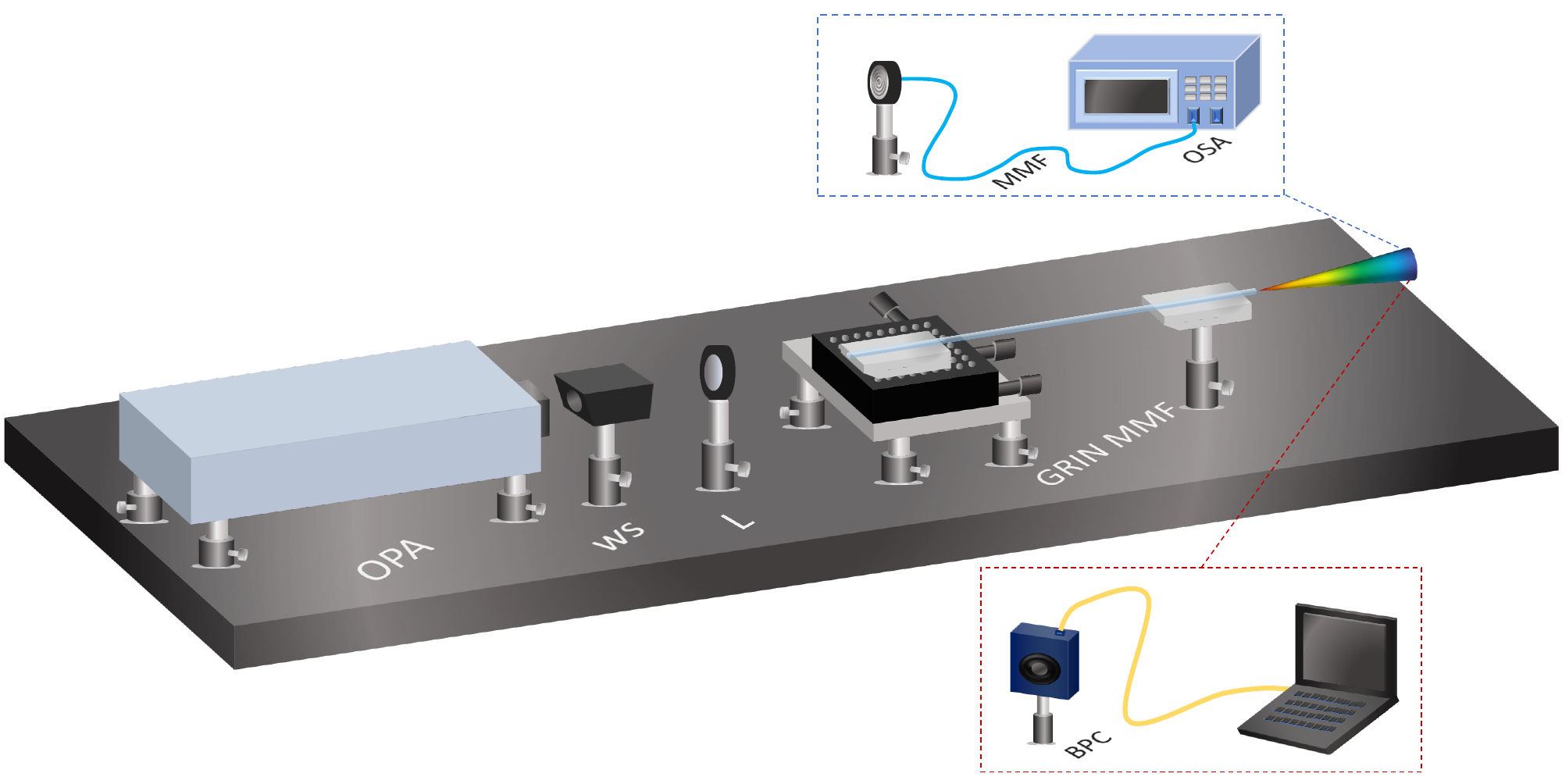}
\caption{Experimental setup. Optical parametric amplifier (OPA); wavelength separator (WS) to select signal/idler; focusing lens (L); collecting multimode fiber (MMF); optical spectrum analyzer (OSA); beam profile camera (BPC).}
\label{fig:setup}
\end{figure}

Our experimental setup is shown in Fig.~\ref{fig:setup}. We used a tunable optical parametric amplifier (OPA) delivering 350~fs pulses and a peak power up to 2~MW. The spatial intensity distribution of the laser beam is Gaussian with a beam quality factor $M^2 < 1.2$. Depending on the desired pump wavelength, the signal or idler can be selected at the output of OPA using a wavelength separator. The 15 cm long tellurite GRIN fiber was placed on a three-axis precision translation stage to control the excitation conditions and maximize the coupling efficiency. The generated SC spectra and corresponding spatial intensity distribution in the far-field were measured as a function of injected pulse energy for a pump wavelength both in the normal and anomalous dispersion regimes. Two different optical spectrum analyzers (OSAs) were used to characterize the SC spectra, covering the spectral ranges from 350 to 1750~nm (Yokogawa-AQ6374) and from 1500 to 3400~nm (Yokogawa-AQ6376), respectively. A beam profile camera (Ophir-Spiricon-Pyrocam IIIHR) was used to record the transverse spatial intensity distribution at the fiber output in the far-field.

\begin{figure*}[!ht]
\centering
\includegraphics[width=0.95\linewidth]{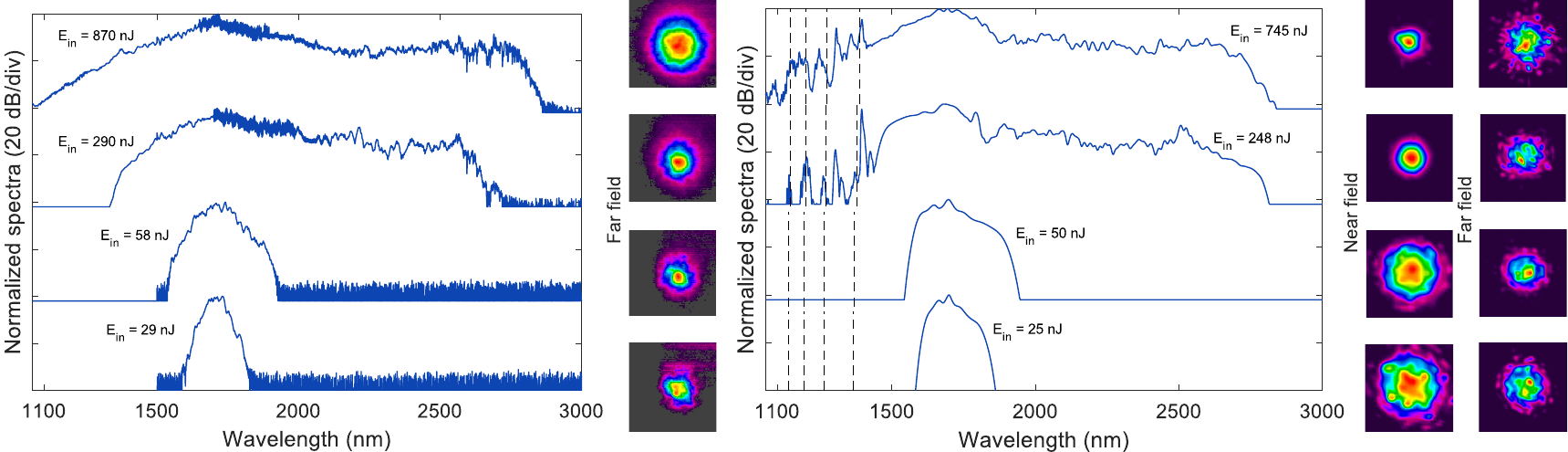}
\caption{Normal dispersion regime with pump wavelength at 1700~nm. Left panel: experimental results of the generated SC and transverse intensity distribution in the far-field vs. injected pulse energy. Right panel: numerical simulations of the SC spectrum and transverse intensity distribution in the near- and far-field vs. injected pulse energy. The dashed lines correspond to the theoretical position of GPI sidebands. The input pulse energies for both experimental results and numerical simulations are indicated in the Figure. Note, that the input energy used in the numerical simulations was adjusted to obtain similar output energy as in the experiments.}
\label{fig:normal}
\end{figure*}

We first tuned the pump wavelength to 1700~nm in the normal dispersion regime. The pulses were injected into the fiber using a plano-convex $\rm{MgF_2}$ lens with a focal length of 50~mm. The beam size at the fiber input was estimated to be 49~µm. Note that the numerical aperture of the injected beam exceeds that of the fiber such that a fraction of the energy is initially coupled into higher-order modes. The throughput was calculated as the ratio of output-to-input energy and measured to be 37~\%. The SC spectrum and spatial intensity profile at the fiber output are shown in Fig.~\ref{fig:normal} for increasing injected pulse energy. The initial mechanism seeding the SC generation process is self-phase modulation (SPM) which manifests as near-symmetrical spectral broadening. 

At larger energies, the spectrum broadens into the anomalous dispersion regime, seeding the generation of solitons. With further increase of energy, soliton self-frequency shift leads to strong spectral asymmetry and more pronounced broadening towards the longer wavelengths. The far-field spatial intensity distribution associated with the SC spectra is also shown in Fig.~\ref{fig:normal}. Theoretically, the fiber supports several hundreds of transverse modes at 1700~nm. When light is injected at normal incidence, the measured spatial intensity distribution at the fiber output becomes less speckled and more uniform as the injected energy is increased, exhibiting a near-Gaussian profile at the highest input energy value. This behavior is consistent with self-cleaning dynamics, however, here, it is important to note that there is significant dissipation due to the large attenuation of the fiber beyond 2900~nm. Also, because the Raman frequency shift plays a significant role in extending the spectrum towards the longer wavelengths, it is possible that the observed cleaning arises from the Raman gain and one cannot therefore attribute with certainty the cleaning dynamics solely to the Kerr effect.

To confirm our experimental observations, we performed numerical simulations of the (3+1)D GNLSE with details available in the Supplementary Document. The numerical simulations shown in Fig.~\ref{fig:normal} (right panel) agree well with the experimental measurements (left panel). The initial spectral broadening is induced by SPM and, at higher input energies, the SC spectrum becomes highly asymmetric towards the long wavelengths due to soliton dynamics. The simulations show the presence of discrete spectral components on the short wavelengths side arising from geometric parametric instabilities (GPI) phase-matched by the periodic beam focusing and defocusing. Theoretically predicted GPI sidebands are marked in the Figure by the dashed lines and they align well with the position in the numerical simulations. However in the experiments, one does not observe clear distinct sidebands which we attribute to variations in the core size along the fiber as mentioned above, preventing efficient phase-matching from being fulfilled. The simulated spatial intensity profile at the fiber output follows similar qualitative behavior as in the experiments, with smoother intensity distribution for increased input energy. 

\begin{figure*}[!ht]
\centering
\includegraphics[width=0.95\linewidth]{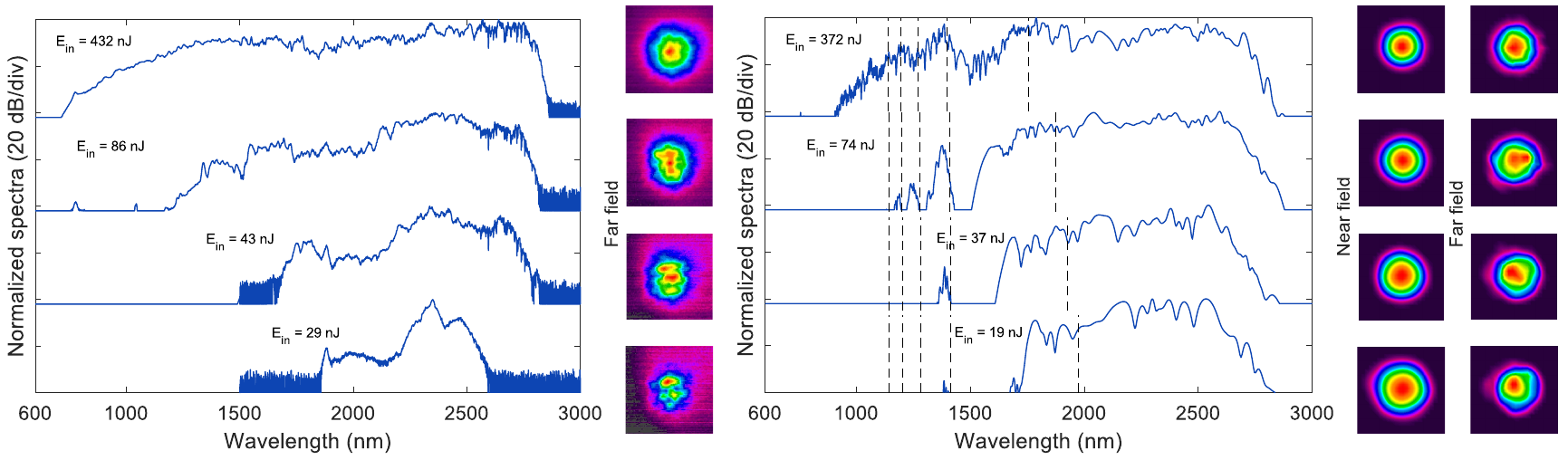}
\caption{Anomalous dispersion regime with pumping wavelength at 2300~nm. Left panel: experimental SC spectrum and transverse intensity distribution in the far-field vs. injected pulse energy. Right panel: numerically simulated SC spectrum and transverse intensity distribution in the near- and far-field vs. injected pulse energy. The dashed lines correspond to the theoretical position of the DWs. The input pulse energies for both experimental results and numerical simulations are indicated in the Figure. Note, that the input energy used in the numerical simulations was adjusted to obtain similar output energy as in the experiments.}
\label{fig:anomalous}
\end{figure*}

We next tuned the pump wavelength to 2300~nm, in the anomalous dispersion regime of the fiber. A plano-convex Si lens with a 25~mm focal length was used in this case to couple light into the fiber. The beam size at the fiber input was estimated to be 33~µm. The results are illustrated in Fig.~\ref{fig:anomalous}. The SC generation process in this regime is initiated by higher-order soliton dynamics: soliton compression leading to near-symmetrical spectral expansion followed by fission and the generation of multiple dispersive waves (DWs) in the normal dispersion phase-matched by the self-imaging phenomenon. The theoretical positions of the DW spectral components shown as dashed lines in the Figure are in good agreement with those in the simulations. This is followed by soliton self-frequency shift expanding the spectrum towards the longer wavelengths and interactions with the DW via cross-phase modulation resulting in a continuous SC spectrum from 790~nm to the edge of the transmission window of the fiber at 2900~nm. Note that since the pump wavelength is relatively close to the edge of the transmission window of the fiber, the spectral broadening on the long wavelengths side saturates at high input energy and the throughput efficiency drops to 9~\% as compared to when the pump wavelength is in the normal dispersion regime. Numerical simulations for the anomalous dispersion pumping regime are shown on the right panel in Fig.~\ref{fig:anomalous}. One observes good correspondence with the experimental results. In particular, the spectral broadening as a function of the input pulse energy is well reproduced and the simulations also clearly display characteristics of multiple DW generation phase-matched by spatial self-imaging. Similarly to the GPI sidebands in the normal regime, the DW components are not as apparent as in the simulations due to the fiber structural inhomogeneity.

The spatial intensity distribution was characterized as a function of input pulse energy and the results are shown in Fig.~\ref{fig:anomalous}. One can see that at low input energy, the output beam profile exhibits a clear speckle pattern indicative of multimode excitation. As energy is increased, the beam intensity profile becomes more symmetric and smooth. This behavior would be expected from Kerr self-cleaning dynamics, but here, the cleanup process may rather be associated with dissipation and in particular Raman dynamics. Numerical simulations on the other hand show that the beam-cleaning process already occurs at low input pulse energies. This may happen since the simulations assume a perfectly homogeneous fiber structure yielding more efficient cleaning dynamics. 

Supercontinuum generation in graded-index multimode fibers has recently attracted significant interest due to the possibility of increasing the power throughput while maintaining a high beam quality. While all previous works have been limited to silica multimode fibers, a new approach based on nanostructuring of multi-component glasses was recently introduced, allowing for the design of soft-glass GRIN fibers with extended transmission windows and high nonlinearity. Here, we have shown that the concept of nanostructuring is generic and can also be applied to tellurite glasses to obtain a GRIN multimode fiber with five times larger nonlinearity compared to that of the PBG GRIN fiber reported earlier, enabling SC generation with comparable bandwidth at much lower input energy. We have conducted a systematic analysis of the generated SC for the normal and anomalous dispersion regimes and characterized the beam intensity profile in the far field. It was shown that soliton dynamics results in a SC spectrum spanning nearly two octaves from 790~nm to 2900~nm when pumping in the anomalous dispersion region, which, to the best of our knowledge, represents the longest wavelength ever reached in graded-index multimode fibers. Also, we observed that at high injected pulse energy the spatial intensity distribution shows signatures of self-cleaning dynamics. Additionally, we performed numerical simulations that are in good agreement with our experimental results. 

The current work shows that multimode graded-index fibers have the potential to overcome the power limitation of SC generation in the mid-infrared electromagnetic region while maintaining a good beam quality. Improvements in the manufacturing process could extend the transmission window further, such that graded-index tellurite fiber is an excellent candidate for SC extension to mid-infrared with many potential applications including LIDAR, bioimaging, and molecular fingerprinting.

\noindent\textbf{Funding.} E.K., L.S., and G.G. acknowledge support from the Academy of Finland (Grants 333949, Flagship PREIN 346511). T.K., M.K., and R.B. acknowledge support from the Foundation for Polish Science co-financed by the European Union under the European Regional Development Fund (Project POIR.04.04.00-00-1644/18).

\noindent\textbf{Disclosures.} 
The authors declare no conflicts of interest.

\noindent\textbf{Data availability.} 
Data are available from the corresponding author upon reasonable request.

\noindent\textbf{Supplementary Document.} 
See the Supplementary Document for supporting content.

\bibliographystyle{ieeetr}
\bibliography{main}

\begin{thebibliography}{10}

\bibitem{article1}
K.~Krupa, C.~Louot, V.~Couderc, M.~Fabert, R.~Guenard, B.~M. Shalaby, A.~Tonello, D.~Pagnoux, P.~Leproux, A.~Bendahmane, R.~Dupiol, G.~Millot, and S.~Wabnitz, ``Spatiotemporal characterization of supercontinuum extending from the visible to the mid-infrared in a multimode graded-index optical fiber,'' {\em Optics Letters}, vol.~41, no.~24, pp.~5785--5788, 2016.

\bibitem{Saini:15}
T.~S. Saini, A.~Kumar, and R.~K. Sinha, ``Broadband mid-infrared supercontinuum spectra spanning 2--15 $\mu$m using {As$_{2}$Se$_3$} chalcogenide glass triangular-core graded-index photonic crystal fiber,'' {\em Journal of Lightwave Technology}, vol.~33, pp.~3914--3920, Sep 2015.

\bibitem{article2}
U.~Te{\u{g}}in and B.~Orta{\c{c}}, ``Cascaded raman scattering based high power octave-spanning supercontinuum generation in graded-index multimode fibers,'' {\em Scientific Reports}, vol.~8, no.~1, p.~12470, 2018.

\bibitem{gpi}
K.~Krupa, A.~Tonello, A.~Barth\'el\'emy, V.~Couderc, B.~M. Shalaby, A.~Bendahmane, G.~Millot, and S.~Wabnitz, ``Observation of geometric parametric instability induced by the periodic spatial self-imaging of multimode waves,'' {\em Physical Review Letters}, vol.~116, p.~183901, 2016.

\bibitem{multimodesoliton}
W.~H. Renninger and F.~W. Wise, ``Optical solitons in graded-index multimode fibres,'' {\em Nature Communications}, vol.~4, no.~1, p.~1719, 2013.

\bibitem{krupanature}
K.~Krupa, A.~Tonello, B.~M. Shalaby, M.~Fabert, A.~Barth{\'e}l{\'e}my, G.~Millot, S.~Wabnitz, and V.~Couderc, ``Spatial beam self-cleaning in multimode fibres,'' {\em Nature Photonics}, vol.~11, no.~4, pp.~237--241, 2017.

\bibitem{scsoftglassreview}
T.~S. Saini and R.~K. Sinha, ``Mid-infrared supercontinuum generation in soft-glass specialty optical fibers: A review,'' {\em Progress in Quantum Electronics}, vol.~78, p.~100342, 2021.

\bibitem{stackanddraw}
D.~Pysz, I.~Kujawa, R.~Stepien, M.~Klimczak, A.~Filipkowski, M.~Franczyk, L.~Kociszewski, J.~Buzniak, K.~Harasny, and R.~Buczynski, ``Stack and draw fabrication of soft glass microstructured fiber optics,'' {\em Bull. Pol. Acad. Sci.-Tech. Sci.}, vol.~62, p.~667, 12 2014.

\bibitem{article8}
R.~Buczynski, ``Photonic crystal fibers,'' {\em Acta Physica Polonica Series B}, vol.~106, pp.~141--166, 08 2004.

\bibitem{Eslami}
Z.~Eslami, L.~Salmela, A.~Filipkowski, D.~Pysz, M.~Klimczak, R.~Buczyński, J.~Dudely~M., and G.~Goëry, ``Two octave supercontinuum generation in a non-silica graded-index multimode fiber,'' {\em Nature Communications}, vol.~13, p.~2126, 2022.

\bibitem{telluritechar}
X.~Feng, P.~Horak, and F.~Poletti, ``Tellurite glass fibers for mid-infrared nonlinear applications,'' {\em Technological Advances in Tellurite Glasses: Properties, Processing, and Applications}, pp.~213--239, 2017.

\bibitem{liao2009tellurite}
M.~Liao, C.~Chaudhari, G.~Qin, X.~Yan, T.~Suzuki, and Y.~Ohishi, ``Tellurite microstructure fibers with small hexagonal core for supercontinuum generation,'' {\em Optics Express}, vol.~17, no.~14, pp.~12174--12182, 2009.

\bibitem{shi2016multi}
H.~Shi, X.~Feng, F.~Tan, P.~Wang, and P.~Wang, ``Multi-watt mid-infrared supercontinuum generated from a dehydrated large-core tellurite glass fiber,'' {\em Optical Materials Express}, vol.~6, no.~12, pp.~3967--3976, 2016.

\bibitem{Saini:19}
T.~S. Saini, N.~P.~T. Hoa, T.~H. Tuan, X.~Luo, T.~Suzuki, and Y.~Ohishi, ``Tapered tellurite step-index optical fiber for coherent near-to-mid-ir supercontinuum generation: experiment and modeling,'' {\em Applied Optics}, vol.~58, no.~2, pp.~415--421, 2019.

\end{thebibliography}

\clearpage

\end{document}